\documentclass[preprint,showpacs,preprintnumbers,amsmath,amssymb]{revtex4-1}

\usepackage{graphicx}
\usepackage{color}
\usepackage{comment}
\usepackage[normalem]{ulem}
\usepackage{bm}

\renewcommand{\bm}[1]{{\mbox{\boldmath $#1$}}}
\newcommand{\CVS}[1]{CsV${_3}$Sb${_5}$}

\begin{document}

\title{Bulk evidence of anisotropic $s$-wave pairing with no sign change in the kagome superconductor CsV$_3$Sb$_5$}
\author{M.~Roppongi$^1$}
\author{K.~Ishihara$^1$}
\author{Y.~Tanaka$^1$}
\author{K.~Ogawa$^1$}
\author{K.~Okada$^1$}
\author{S.~Liu$^1$}
\author{K.~Mukasa$^1$}
\author{Y.~Mizukami$^{1,{\S}}$}
\author{Y.~Uwatoko$^2$}
\author{R.~Grasset$^3$}
\author{M.~Konczykowski$^3$}
\author{B.~R.~Ortiz$^4$}
\author{S.~D.~Wilson$^4$}
\author{K.~Hashimoto$^1$}\email{k.hashimoto@edu.k.u-tokyo.ac.jp}
\author{T.~Shibauchi$^1$}\email{shibauchi@k.u-tokyo.ac.jp}
\affiliation{
$^1$Department of Advanced Materials Science, University of Tokyo, Kashiwa, Chiba 277-8561, Japan\\
$^2$Institute for Solid State Physics, University of Tokyo, Kashiwa, Chiba, 277-8581, Japann\\
$^3$Laboratoire des Solides Irradi{\' e}s, CEA/DRF/IRAMIS, Ecole Polytechnique, CNRS, Institut Polytechnique de Paris, F-91128 Palaiseau, France\\
$^4$Materials Department, University of California Santa Barbara, Santa Barbara, California 93106, USA\\
$^{\S}$\rm{Present address: Department of Physics, Tohoku University, Sendai 980-8578, Japan.}
}
\date{\today}

\begin{abstract}
{\bf
The recently discovered kagome superconductors \bm{A}V$\bm{_3}$Sb$\bm{_5}$ (\bm{A} = K, Rb, Cs) \,\cite{Ortiz_PRL,RbVS_SC,Ortiz_PRMat_2021} possess a unique band structure with van Hove singularities and Dirac dispersions\,\cite{Ortiz_PRX,kang_nat_phys_FS}, in which unusual charge-density-wave (CDW) orders with time-reversal and rotational symmetry breaking have been reported\,\cite{Hasan_CDW_natmat,CVS_nature_cascade,KVS_muSR_nature,CVS_nature_nematic,Nematic_KVS_natphys,Nem_NatCom}. One of the most crucial unresolved issues is identifying the symmetry of the superconductivity that develops inside the CDW phase. Theory predicts a variety of unconventional superconducting symmetries, including exotic states with chiral and topological properties accompanied by a sign-changing superconducting gap\,\cite{Kiesel_PRB_2012, Kiesel_PRL_2013,Wang_PRB_2013,Thomale_PRL_2021,tazai_kagome}. Experimentally, however, the phase information on the superconducting gap in \bm{A}V$\bm{_3}$Sb$\bm{_5}$ is still lacking. Here we report the electron irradiation effects in $\bm{\rm CsV{_3}Sb{_5}}$ using introduced impurities as a phase-sensitive probe of superconductivity. Our magnetic penetration depth measurements reveal that with increasing impurities, a highly anisotropic fully-gapped state changes gradually to an isotropic full-gap state without passing through a nodal state. Furthermore, transport measurements under high pressure show that the double superconducting dome in the pressure-temperature phase diagram survives against sufficient impurities. These results are strong bulk evidence that $\bm{\rm CsV{_3}Sb{_5}}$ is a non-chiral, anisotropic $s$-wave superconductor with no sign change both at ambient and high pressure, which provides a clue to understanding the relationship between CDW and superconductivity in kagome superconductors.}
\end{abstract}

\clearpage

\maketitle


The kagome lattice, a motif consisting of corner-sharing triangles and hexagonal holes, provides a platform for a rich variety of novel quantum phases of matter. Due to its strong geometrical frustration, it has long been studied in quantum spin systems as a playground for realizing quantum spin liquids\,\cite{Herbertsmithite_review}. Recently, however, significant efforts have been devoted to exploring topological metals and semimetals in kagome-lattice systems, in which unique band structures such as flat bands, Dirac cones, and van Hove singularities (vHSs) can lead to Dirac/Weyl fermions\,\cite{FeSn_nature,Mn3Sn_nature_2015}, spin/charge ordering, and unconventional superconductivity\,\cite{Kiesel_PRB_2012, Wang_PRB_2013, Kiesel_PRL_2013}. Although various topological kagome materials have been reported so far\,\cite{FeSn_nature,Mn3Sn_nature_2015}, superconductors with kagome lattices are rarely found\,\cite{MOF_takenaka}.

The recently discovered $A$V$_3$Sb$_5$ ($A$ = K, Rb, Cs) is a new family of kagome superconductors with the superconducting transition temperature $T_{\rm c}$ of 0.9--2.5\,K \cite{Ortiz_PRL,RbVS_SC,Ortiz_PRMat_2021}. The alkali $A$ atoms are intercalated between sheets consisting of the two-dimensional (2D) kagome networks of V atoms and triangular and hexagonal networks of Sb atoms (Fig.\,\ref{F1}a,b). The electronic band dispersions near the Fermi energy $E_{\mathrm{F}}$ share characteristic features predicted for an ideal kagome-lattice system such as a vHS at the M point and a Dirac point at the K point\,\cite{Ortiz_PRMat_2019,Ortiz_PRL,Ortiz_PRMat_2021,kang_nat_phys_FS,Ortiz_PRX}. In these materials, $E_{\mathrm{F}}$ is located near the vHS point, and multiple Fermi surfaces are formed by the V $d$-orbitals and Sb $p$-orbitals (Fig.\,\ref{F1}c).
Such unique band structures in $A$V$_3$Sb$_5$ give rise to unconventional charge-density-wave (CDW) orders with the transition temperature $T^{\ast} \sim 78 $--$103$\,K\,\cite{Ortiz_PRL, Ortiz_PRMat_2021, RbVS_SC} driven by electron correlation\,\cite{phonon_cdw, Wang_PRB_2013, Thomale_PRL_2021, Balents_PRB_2021, Fernandes_PRB_2021}. The CDW transition is accompanied by a $2a_0 \times 2a_0 \times 2c_0$ or $2a_0 \times 2a_0 \times 4c_0$ superlattice composed of modulated star-of-David and inverse star-of-David patterns (where $a_0$ and $c_0$ indicate the lattice constants above $T^{\ast}$), which breaks translational symmetry\,\cite{Hasan_CDW_natmat,Ortiz_PRX,Hu_2022_SOD,Kang_2022_SOD}. More intriguingly, it has been reported that additional symmetries, such as time-reversal symmetry (TRS) and rotational symmetry (RS), can be broken below $T^{\ast}$\,\cite{Hasan_CDW_natmat,CVS_nature_cascade,KVS_muSR_nature,CVS_nature_nematic,Nematic_KVS_natphys,Nem_NatCom}. Since the superconducting transition takes place inside the unusual CDW phase, a fundamental question arises as to whether the superconducting pairing mechanism in $A$V$_3$Sb$_5$ is conventional or not\,\cite{Kagome_review_natphys}.

Theories on the kagome lattice near van Hove filling have proposed that unconventional superconductivity beyond the electron-phonon mechanism can be realized by electron correlation effects\,\cite{Kiesel_PRB_2012, Kiesel_PRL_2013, Wang_PRB_2013, Thomale_PRL_2021, tazai_kagome}. Spin and charge fluctuations can lead to spin-triplet $p$- and $f$-wave\,\cite{Kiesel_PRL_2013, Thomale_PRL_2021, tazai_kagome} and chiral $d$-wave superconductivity\,\cite{Thomale_PRL_2021}, whereas bond-order fluctuations can promote anisotropic $s$-wave\,\cite{tazai_kagome} and chiral $d$-wave superconductivity\,\cite{Wang_PRB_2013}. In support of the above, first-principle calculations have pointed out that the Bardeen-Cooper-Schrieffer (BCS) theory (electron-phonon mechanism) cannot explain the experimental $T_{\rm c}$ values, suggesting an unconventional pairing mechanism in $A$V$_3$Sb$_5$\,\cite{AVS_first_pri_PRL}.

Experimentally, however, the superconducting gap symmetry of $A$V$_3$Sb$_5$ is highly controversial, and whether TRS is broken or not is still elusive. Thermal conductivity measurements in \CVS\,\,\cite{SYLi_thermal} and $\mu$SR measurements in Rb/KV$_3$Sb$_5$\,\cite{CVS_muSR_SC_ambient} have suggested a nodal gap structure. In contrast, magnetic penetration depth\,\cite{CVS_TDO_Yuan} and scanning tunneling spectroscopy (STS)\,\cite{CVS_STM_imp_PRL} studies in \CVS\, have suggested a nodeless gap structure. Nuclear magnetic/quadrupole resonance (NMR/NQR) measurements in \CVS\,\,\cite{CVS_NMR} have shown a finite Hebel-Slichter coherence peak in $1/T_1T$ and a decrease in Knight shift below $T_{\rm{c}}$, which exclude spin-triplet superconductivity. Regarding the TRS in the superconducting state, Josephson STS measurements in \CVS\,\,\cite{Roton_PDW_nature} have suggested a possible roton pair-density wave, corresponding to an unconventional superconducting state with TRS breaking. Contrastingly, $\mu$SR studies in \CVS\,\,\cite{CVS_muSR_SC_ambient} have reported that TRS is not broken in the superconductivity state.
In addition to the above results at ambient pressure, high-pressure studies\,\cite{CVS_pressure_PRL, CVS_pressure_ncomm} have revealed that the CDW phase is suppressed by the application of pressure, accompanied by the emergence of a superconducting dome, indicating the close relationship between the CDW and superconductivity. Moreover, recent $\mu$SR experiments under pressure\,\cite{CVS_muSR_pressure, RbVS_muSR_pressure} have suggested that TRS is broken in the superconducting state when the CDW phase is suppressed by applying pressure. Therefore, to clarify the pairing mechanism of the kagome superconductors, it is crucial to pin down the superconducting gap symmetry of $A$V$_3$Sb$_5$ both at ambient and high pressure, including whether TRS is broken or not.

In general, the conventional phonon-mediated pairing mechanism leads to a superconducting gap opening all over the Fermi surface, while unconventional pairing mechanisms, such as spin fluctuations, can lead to an anisotropic gap with nodes where the superconducting gap becomes zero. Thus, experimental observations of the low-energy quasiparticle excitations determine whether the gap structure has nodes or not. In addition to clarifying the presence or absence of nodes in the gap, determining the sign of the gap function also provides a strong constraint on the superconducting pairing symmetry. Especially in kagome superconductors, theories have predicted that two degenerated superconducting order parameters, $d_{x^2-y^2}$ and $d_{xy}$, give rise to chiral $d_{x^2-y^2}\pm id_{xy}$-wave symmetry, where a finite gap opens all over the Fermi surface, but the phase of the order parameter changes by $4\pi$ in momentum $\bm{k}$ space\,\cite{Yonezawa}. Therefore, phase-sensitive probes are highly required to determine the pairing symmetry of $A$V$_3$Sb$_5$.

\begin{figure*}[tbp]
    \includegraphics[width=\linewidth]{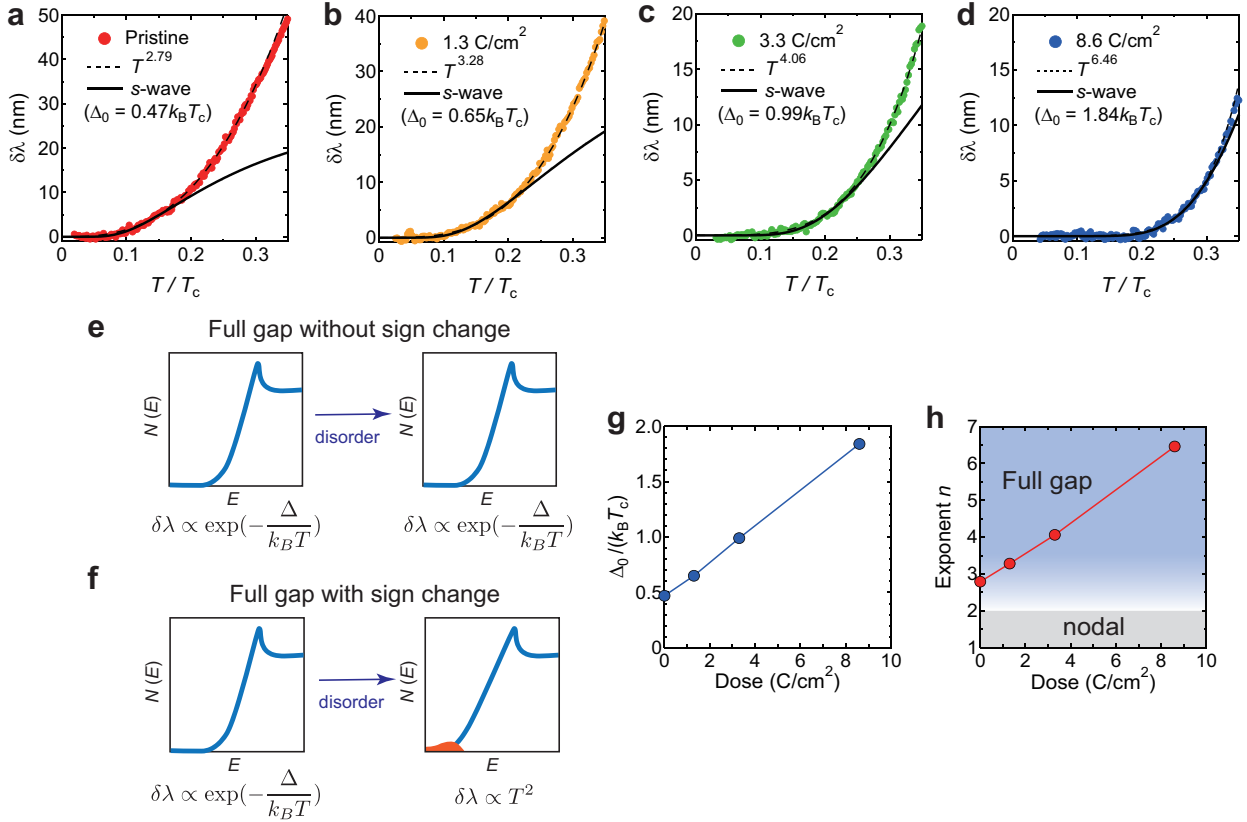}
    \clearpage
    \caption{
       {\bf Electron irradiation effects on the low-temperature penetration depth.} {\bf a}-{\bf d}, Temperature dependence of the change in the penetration depth $\delta \lambda (T)$ for the pristine\,({\bf a}), 1.3\,({\bf b}), 3.3\,({\bf c}), and 8.6\,({\bf d}) C/cm$^2$ irradiated samples. Solid circles are the experimental data. Black solid and dashed lines are fitting curves of the fully-gapped $s$-wave and power-law analysis. {\bf e},{\bf f}, Schematics of the change in the quasiparticle density of states against disorder in the case of fully gapped states without ({\bf e}) and with ({\bf f}) sign change. {\bf g}, Gap value $\Delta_{\rm 0}/\,k_{\rm B}T_{\rm c}$ obtained from the $s$-wave fit to $\delta \lambda(T)$ at low tempeartures as a function of irradiation dose. {\bf h}, Exponent value $n$ obtained from the power-law fitting to $\delta \lambda(T)$ up to $0.3\,T_{\rm c}$ as a function of irradiation dose. The exponent $n \leq 2$ (gray shaded region) indicates a nodal gap structure, while $n \gtrsim 3$ implies a fully gapped state (blue shaded region). }
    \label{F2}
\end{figure*}

There are several experimental probes that are sensitive to the sign of gap functions, such as Josephson junction\,\cite{cuprate_junction}, quasiparticle interference\,\cite{hanaguri_science}, and neutron scattering techniques\,\cite{BaK122_neutron}. In general, however, the analysis of such interference effects is complicated in multiband systems due to the complexity of the scattering processes. In addition, most of them require good surface/interface conditions. In contrast, the non-magnetic impurity effect, on which we focus here, is one of the phase-sensitive probes that are applicable to multiband systems and reflect the bulk superconducting properties\,\cite{mizukami_Ba122_ncom,BaRu122_irr_prozorov}. In $s$-wave superconductors with no sign-changing order parameter, the Cooper pairs are not destroyed by non-magnetic impurity scattering, and both $T_{\rm c}$ and quasiparticle density of states (DOS) are little affected by disorder (the so-called Anderson's theorem\,\cite{anderson_dirty}) (Fig.\,\ref{F2}e). In contrast, in the case of nodeless superconductors with sign-changing order parameters, such as chiral $d$-wave and $s_{\pm}$-wave superconductors, the Cooper pairs are destroyed by impurity scattering, which suppresses $T_{\rm c}$ rapidly and induces impurity states associated with the Andreev bound state \,\cite{mizukami_Ba122_ncom,BaRu122_irr_prozorov} (Fig.\,\ref{F2}f). In this case, additional low-energy quasiparticle excitations appear near the zero energy, e.g., leading to a change in the temperature dependence of the magnetic penetration depth $\lambda$ from exponential to $T^2$. 

In this study, we used electron irradiation to systematically introduce non-magnetic impurities into \CVS\, single crystals (see Methods and Supplementary Information Sec.\,$\mathrm{I}$). In this method, high-energy electron beam irradiation creates vacancies in the crystal\,\cite{mizukami_Ba122_ncom}, acting as point defects without changing the electronic structure and lattice constants (see Supplementary Information Sec.\,$\mathrm{II}$). Figure\,\ref{F1}d,e shows the temperature dependence of resistivity $\rho (T)$ at ambient pressure in samples with irradiated doses of 0 (pristine), 1.3, 3.3, and 8.6\,C/cm$^2$. The residual resistivity ratio (RRR) of the pristine sample is $\sim$\,84, indicating the high quality of our crystals. As the dose increases, the residual resistivity $\rho_0$ increases (also see Fig.\,\ref{F1}h), and the RRR value decreases. Along with this, both the CDW and superconducting transition temperatures $T^{\ast}$ and $T_{\rm c}$ shift to a lower temperature (see Fig.\,\ref{F1}g). The suppression of $T_{\rm c}$ has also been confirmed by the Meissner effect measured by the normalized frequency shift of a tunnel diode oscillator (TDO) (see Fig.\,\ref{F1}f). Note that the superconducting transition becomes sharper with increasing dose, which may be related to the suppression of superconducting phase fluctuations\,\cite{MOF_takenaka,Emery1995} or the change in skin depth due to impurity scattering. The sharp superconducting transition width in the 8.6\,C/cm$^2$ irradiated sample with sufficient disorder indicates that the defects are introduced quite uniformly inside the crystals.

Next, we turn to the impurity effect on low-energy quasiparticle excitations in the superconducting state. Magnetic penetration depth $\lambda$ is one of the most fundamental properties of superconductors sensitive to low-energy quasiparticle excitations\,\cite{CVS_TDO_Yuan,MOF_takenaka,mizukami_Ba122_ncom}. In this study, we measured the magnetic penetration depth of the pristine and irradiated \CVS\, single crystals down to 50\,mK by using the TDO in a dilution refrigerator (see Methods). Figure\,\ref{F2}a-d shows the change in the magnetic penetration depth $\delta \lambda(T) \equiv \lambda(T)-\lambda(0)$ (where $\lambda(0)$ is the absolute value of the penetration depth at 0\,K) at low temperatures for the pristine and 1.3, 3.3, and 8.6\,C/cm$^2$ irradiated samples. 
In the pristine sample, $\delta \lambda (T)$ shows a flat temperature dependence at low temperatures below 0.1$T_{\rm c}$ (Fig.\,\ref{F2}a), indicating a fully-gapped superconducting state in \CVS\,. To examine the low-energy quasiparticle excitations in the pristine sample, we applied a power-law fit $\delta\lambda(T) \propto T^{n}$ to the experimental data. In general, in the case of nodal superconductors with line and point nodes, the exponent value $n$ gives 1 and 2 in the clean limit, respectively. We obtained $n \sim\,2.8$ from the fitting (Fig.\,\ref{F2}a), indicating the absence of nodes in the gap (or conversely, the presence of a finite gap). Then, to quantitatively evaluate the gap value, we tried to fit the data with the fully-gapped $s$-wave model $\delta\lambda(T)\propto T^{-1/2} \exp(-\Delta_{0}/k_{\rm B}T)$, where $k_{\rm B}$ is the Boltzmann constant and $\Delta_{0}$ is the superconducting gap. We obtained an extremely small gap value $\Delta_{0}=0.47k_{\rm B}T_{\rm c}$ (which is consistent with the previous study\,\cite{CVS_TDO_Yuan}), suggesting the existence of gap minima $\Delta_{\rm min}$ coming from the anisotropic gap nature of \CVS\,, as discussed later.
One of our key findings is that the fully-gapped behavior in $\delta\lambda(T)$ is robust against disorder (see Fig.\,\ref{F2}b-d). The flat temperature region at low temperatures expands to a higher temperature region with increasing dose. In the case of fully-gapped superconductors with sign-changing order parameters, $\delta \lambda(T)$ is expected to change from an exponential to a $T^2$ dependence with increasing impurities because of the impurity-induced DOS (see Fig.\,\ref{F2}f)\,\cite{mizukami_Ba122_ncom}. In sharp contrast, our experimental observations show that $\Delta_{\rm 0}$ and $n$ obtained from the fitting rather increase with increasing dose (Fig.\,\ref{F2}g,h), indicating no impurity-induced DOS in the superconducting gap.
These results provide strong bulk evidence that the superconducting gap structure of \CVS\, is nodeless without a sign-changing gap.

For a more detailed analysis of the superconducting gap structure, we derived the normalized superfluid density $\rho_{\rm s}(T) \equiv  \lambda^2(0)/\lambda^2(T)$. We used $\lambda(0)=387$\,nm for the pristine sample estimated in the previous study\,\cite{CVS_TDO_Yuan} and calculated $\lambda(0)$ for the irradiated samples by using the relation $\lambda(0)=\lambda_{\rm L}(0)(1+\xi/l)^{1/2}$ (see Fig.\,\ref{F1}h), where the London penetration depth $\lambda_{\rm L}(0)$ is assumed to be equal to $\lambda(0)=387$\,nm for the pristine sample, and $l$ and $\xi$ are the mean free path and coherence length, respectively (for more details, see Supplementary Information Sec.\,$\mathrm{IV}$). Figure\,\ref{F3}a shows the obtained $\rho_{\rm s}(T)$ curve as a function of $T/T_{\rm c}$ for each sample. In all the samples, $\rho_{\rm s}$ shows a flat temperature dependence at low temperatures, which extends to a higher temperature region with increasing dose. This is again inconsistent with a nodal gap structure. 

Here, we consider a multigap model to analyze the overall temperature dependence of $\rho_{\rm s}$. In \CVS\,, the Fermi surfaces are formed by two different orbitals: one is derived from the $d$-orbitals of V, forming a hexagonal Fermi surface around the $\Gamma$ point and two triangular Fermi surfaces around the K point, while the other is from the $p$-orbitals of Sb, forming a circular Fermi surface around the $\Gamma$ point (see Fig.\,\ref{F1}c)\,\cite{Ortiz_PRX}. The Fermi surfaces derived from the V $d$-orbitals determine the physical properties of this material, and three equivalent $\bm{q}$ vectors\,\cite{Hasan_CDW_natmat,CVS_nature_cascade,CVS_nature_nematic,Nematic_KVS_natphys} are considered to give rise to anisotropic pairing interactions\,\cite{Wang_PRB_2013,tazai_kagome}. Indeed, recent STM measurements\,\cite{CVS_STM_imp_PRL} have reported the emergence of an anisotropic superconducting gap as well as an isotropic gap below $T_{\rm c}$. We therefore consider a multigap model with an anisotropic but nodeless superconducting gap with six-fold symmetry ($\Delta_1 \propto 1+\alpha\,\mathrm{cos}(6\phi)$) and an isotropic superconducting gap ($\Delta_2 = \rm{const.}$) on two cylindrical Fermi surfaces (see Fig.\,\ref{F3}b). We fitted the experimental data with this model (Fig.\,\ref{F3}a) and obtained the gap values $\Delta_1$ and $\Delta_2$ as a function of dose (Fig.\,\ref{F3}c). As the dose increases, the difference between the maximum and minimum values of $\Delta_1$ decreases, and eventually, all the gaps become almost identical.  This is due to the averaging effect between the two gaps introduced by impurity-induced intra/interband scattering, and a very similar behavior has been observed in the prototypical multigap superconductor MgB$_2$\,\cite{MgB2_neutron_irr}. This evidences nodeless multigap superconductivity with a sign-preserving order parameter in \CVS\,, which excludes the possibility of spin-triplet $p$- and $f$-wave and chiral $d$-wave superconductivity.

To discuss the impurity effect on $T_{\rm c}$ more quantitatively, we next introduce a pair-breaking parameter $g=\hbar/(\tau_{\rm imp}k_{\rm B}T_{\rm c0})$, where $\tau_{\rm imp}=\mu_0\lambda_{\rm L}^2(0)/\rho_0$ is the impurity scattering time and $T_{\rm c0}$ is the superconducting transition temperature of the pristine sample\,\cite{BaRu122_irr_prozorov,CCS_irr_yamashita}. The suppression of $T_{\rm c}$ is plotted as a function of $g$ and compared to other superconductors with and without sign-changing order parameters (Fig.\,\ref{F3}d). $T_{\rm c}$ of \CVS\, is rapidly suppressed at a low irradiation dose but starts to saturate at moderate irradiation doses. The initial rapid suppression of $T_{\rm c}$ is considered to be related to the reduction of the anisotropy of $\Delta_{\rm 1}$ (see Fig.\,\ref{F3}c), as discussed later. The $T_{\rm c}$ suppression above 1.3\,C/cm$^2$ irradiation dose is much slower than those in superconductors with sign-changing order parameters such as $d$-wave and $s_{\pm}$-wave superconductors, rather similar to those in $s$-wave superconductors without sign-changing gaps. These results also support that multigap $s$-wave superconductivity with no sign change is realized in CsV$_3$Sb$_5$ at ambient pressure.

\begin{figure*}[tbp]
    \includegraphics[width=\linewidth]{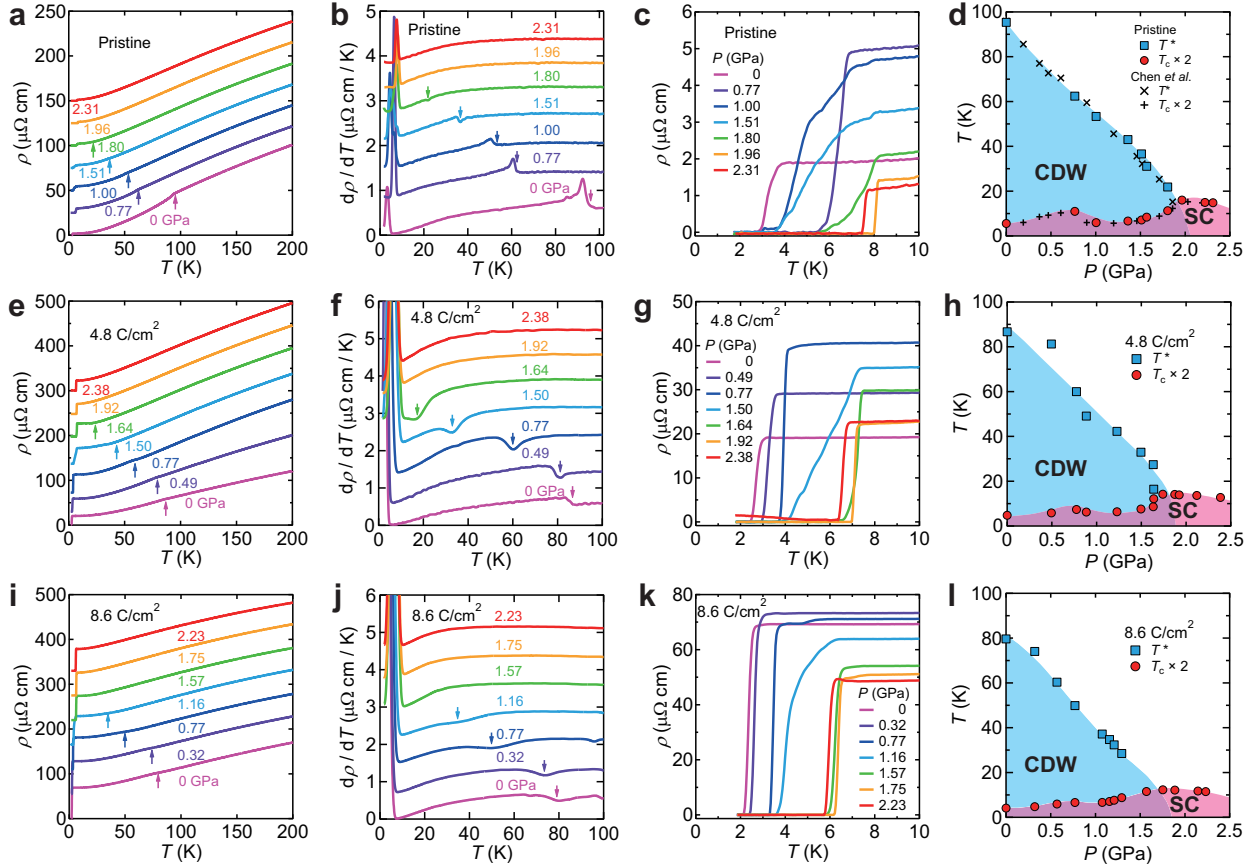}
    \clearpage
    \caption{
{\bf Pressure versus temperature phase diagrams of pristine and irradiated CsV$_3$Sb$_5$.} {\bf a},\,{\bf e},\,{\bf i}, $\rho (T)$ curves below 200\,K at several pressures for the pristine ({\bf a}), 4.8 ({\bf e}), and 8.6 ({\bf i}) C/cm$^2$ irradiated samples. Arrows indicate the CDW transitions determined from the $d\rho/dT$ curves in {\bf b},\,{\bf f},\,{\bf j}. {\bf b},\,{\bf f},\,{\bf j}, Temperature dependence of $d\rho/dT$ below 100\,K for the pristine ({\bf b}), 4.8 ({\bf f}), and 8.6 ({\bf j}) C/cm$^2$ irradiated samples. Arrows indicate the CDW transitions. {\bf c},\,{\bf g},\,{\bf k}, Low-temperature $\rho (T)$ curves below 10\,K at several pressures for the pristine ({\bf c}), 4.8 ({\bf g}), and 8.6 ({\bf k}) C/cm$^2$ irradiated samples. {\bf d},\,{\bf h},\,{\bf l}, $P$-$T$ phase diagrams of the pristine ({\bf d}), 4.8 ({\bf h}), and 8.6 ({\bf l}) C/cm$^2$ irradiated samples. For clarity, $T_{\rm c}$ is doubled. The phase diagram of the pristine sample includes data from Chen $et\,al$\,\cite{CVS_pressure_PRL}.
}
\label{F4}
\end{figure*}

To further investigate the non-magnetic impurity effect on the superconducting phase of \CVS\, under pressure, we constructed the pressure-temperature ($P$-$T$) phase diagram in the pristine sample and 4.8 and 8.6\,C/cm$^2$ irradiated samples. Figure\,\ref{F4}a shows the $\rho(T)$ curve of the pristine sample at several pressures. The CDW transition temperature $T^{\ast}$, which is determined from a jump or dip in $d\rho(T)/dT$ (Fig.\,\ref{F4}b), decreases monotonically with increasing pressure (see Fig.\,\ref{F4}d). 
In contrast, the superconducting transition temperature $T_{\rm c}$ shows a non-monotonic pressure dependence (Fig.\,\ref{F4}c), and a double superconducting dome is observed, as reported in previous high-pressure studies \,\cite{CVS_pressure_PRL, CVS_pressure_ncomm} (Fig.\,\ref{F4}d). The first peak of the superconducting double dome locates at $P_1\sim0.7$\,GPa inside the CDW phase, while the second peak locates at $P_2\sim2$\,GPa near the CDW endpoint. We conducted the same experiments for the 4.8 and 8.6\,C/cm$^2$ irradiated samples (see Fig.\,\ref{F4}e-l) and obtained the $P$-$T$ phase diagrams as shown in Fig.\,\ref{F4}h,l. $T^{\ast}$ is suppressed in the 4.8 and 8.6\,C/cm$^2$ irradiated samples, and the CDW endpoint shifts to lower pressure with increasing irradiation dose. $T_{\rm c}$ is also suppressed by disorder, but the superconducting dome survives even after 8.6\,C/cm$^2$ irradiation. As already discussed in Fig.\,\ref{F3}d, the irradiation dose of 8.6\,C/cm$^2$ introduces enough defects to suppress superconductivity with a sign-changing order parameter. Therefore, these results suggest that the superconducting gap symmetry of \CVS\, at high pressure is also non-sign-changing $s$-wave. Note that recent $\mu$SR measurements under pressure\,\cite{CVS_muSR_pressure} have reported that the superconducting pairing symmetry near $P_2$ has a finite superconducting gap across the Fermi surface and breaks TRS. In such a case, the superconductivity should be sensitive to impurities, inconsistent with our present observations.

Another important aspect of our findings is that the CDW endpoint shifts to lower pressure with irradiation, followed by the second peak of the superconducting double dome (see Fig.\,\ref{F4}d,h,l), suggesting that the CDW is closely related to the superconductivity in the present system. Recent theoretical calculations in $A$V$_3$Sb$_5$\,\cite{tazai_kagome} have proposed that bond-order fluctuations originating from the triple-$\bm{q}$ vectors corresponding to the (inverse) star of David pattern induce anisotropic pairing interactions, leading to anisotropic $s$-wave superconductivity. This theory can explain the relatively high $T_{\rm{c}}$ in $A$V$_3$Sb$_5$ that cannot be reproduced by the BCS theory\,\cite{AVS_first_pri_PRL} and the anisotropic superconducting gap structure in \CVS\, obtained in the present study. Moreover, in such anisotropic $s$-wave superconductivity, the introduction of impurity scattering averages out the anisotropic gap, changing to the isotropic gap. In this case, $T_{\rm{c}}$ drops rapidly at an initial introduction of impurities, but as the gap becomes isotropic, the reduction of $T_{\rm{c}}$ saturates and becomes much slower than that expected in the Abrikosov-Gor'kov (AG) theory. These expectations are in good agreement with our observations of the $T_{\rm{c}}$ suppression in \CVS\,. Thus, our present results support a new type of unconventional superconductivity due to bond-order fluctuations on the kagome lattice in \CVS\,, where the gap function is non-sign-changing $s$-wave. In the present kagome superconductors, the possibility of a loop-current phase with broken TRS and a nematic phase with broken RS has been pointed out above the superconducting phase\,\cite{Hasan_CDW_natmat,CVS_nature_cascade,KVS_muSR_nature,CVS_nature_nematic,Nematic_KVS_natphys,Nem_NatCom}. Therefore, elucidating the intertwining of these unusual normal and superconducting phases, which is commonly seen in high-$T_{\rm c}$ cuprates and iron-based superconductors, will pave the way to understanding novel quantum phases of matter in condensed matter physics.

\clearpage

\noindent
{\bf METHODS}

\noindent
{\bf Single crystal growth.}
High-quality single crystals of CsV$_3$Sb$_5$ were synthesized using the self-flux method. All sample preparations are performed in an argon glovebox with oxygen and moisture $<\,$0.5ppm. The flux precursor was formed through mechanochemical methods by mixing Cs metal (Alfa 99.98\%), V powder (Sigma 99.9\%), and Sb beads (Alfa 99.999\%) to form a mixture which is approximately 50 at.\,\% Cs$_{0.4}$Sb$_{0.6}$ (near eutectic composition) and 50 at.\,\% VSb$_2$. Note that prior to mixing, as-received vanadium powders were purified in-house to remove residual oxides. After milling for 60\,min a pre-seasoned tungsten carbide vial, flux precursors are extracted and sealed into 10\,mL alumina crucibles. The crucibles are nested within stainless steel jackets and sealed under argon. Samples are heated to 1000\,\(^\circ\)C at 250\,\(^\circ\)C/hr and soaked for 24\,h before dropping to 900\,\(^\circ\)C at 100\,\(^\circ\)C/h. Crystals are formed during the final slow cool to 500\,\(^\circ\)C at 1\,\(^\circ\)C/hr before terminating the growth. Once cooled, the crystals are recovered mechanically. Samples are hexagonal flakes with a brilliant metallic luster. The elemental composition of crystals was assessed using energy-dispersive X-ray spectroscopy (EDS) using an APREO-C scanning electron microscope.

\vspace{20pt}
\noindent
{\bf Electron irradiation.}
Electron irradiation with the incident electron energy of 2.5\,MeV was performed on SIRIUS Pelletron linear accelerator operated by the Laboratoire des Solides irradi{\'e}s (LSI) at {\'E}cole Polytechnique. To prevent defect migration and agglomeration, the sample temperature was kept at $\sim\,$20\,K during irradiation which produces stable vacancy-interstitial Frenkel pairs. Subsequent warming to room temperature causes annealing of interstitials, which have a lower migration energy, leaving a uniform population of vacancy type defects. The electron irradiation of 1.3, 3.3, and 8.6\,C/cm$^2$ was performed at the same beam time (run\#1), while the irradiation of 4.8\,C/cm$^2$ was conducted at another beam time (run\#2).

\vspace{20pt}
\noindent
{\bf Electrical resistivity measurements.}
The electrical resistivity was measured at ambient and high pressure by the 4-terminal method using a Physical Property Measurement System (PPMS) from Quantum Design with the lowest temperature of about 1.8\,K. The resistivity under pressure was measured using a piston cylinder cell to generate pressure up to $\sim\,$2.5\,GPa and daphne 7373 as a pressure mediator in PPMS. The pressure value in the sample was determined from the superconducting transition temperature $T_{\rm c}$ of Pb under pressure, using the relation $P=(7.20-T_{\rm c})/0.365$. Note that when the 4.8\,C/cm$^{2}$ irradiated sample was set in the piston cell, even before pressure was applied, the resistivity value changed, probably due to cracks, so we have corrected it to the value before pressure was applied.

\vspace{20pt}
\noindent
{\bf Magnetic penetration depth measurements.}
The temperature variation of the in-plane magnetic penetration depth $\delta \lambda(T) = \lambda(T) - \lambda(0)$ was measured by using a tunnel diode oscillator technique (TDO) with the resonant frequency of $\sim\,$13.8\,MHz in a dilution refrigerator down to $\sim\,$50\,mK. The sample was mounted on a sapphire rod with Apiezon N grease, then inserted into a copper coil in the LC circuit. The frequency shift $\delta f$ in the TDO is related to the change of magnetic susceptibility $\delta\chi$ by the following equation, $\delta f = - (f_0V_{\rm s}/2V_{\rm c}(1-N))\delta\chi$, where $f_0$ is the resonant frequency without the sample, $V_{\rm s}$ and $V_{\rm c}$ are the sample and coil volume, respectively, and $N$ is the demagnetization factor. 
$\delta\chi$ is related to $\delta \lambda$ by the following equation, $\delta\chi = \delta\lambda/R$, where $R$ is a constant determined by the geometry of the sample from the calculation. Thus, $\delta f$ is related to $\delta \lambda$ by the following equation, $\delta f = - (f_0V_{\rm s}/2RV_{\rm c}(1-N))\delta\lambda$.

\bigskip

\noindent{\bf Acknowledgements}\\
\noindent
We thank R. Tazai, Y. Yamakawa, S. Onari, and H. Kontani for fruitful discussions. Electron irradiation was conducted at the SIRIUS accelerator facility at {\'E}cole Polytechnique (Palaiseau, France) and was supported by EMIR\&A French network (FR CNRS 3618) (proposal No. 22-8950). 

\noindent{\bf Funding:} This work was supported by Grants-in-Aid for Scientific Research (KAKENHI) (Nos.\,JP22J21896,\,JP22H00105,\,JP21H01793,\,JP19H00649,\,JP18H05227,\,JP18KK0375), Grant-in-Aid for Scientific Research on innovative areas ``Quantum Liquid Crystals'' (No.\,JP19H05824) and Grant-in-Aid for Scientific Research for Transformative Research Areas (A) ``Condensed Conjugation'' (No.\,JP20H05869) from Japan Society for the Promotion of Science (JSPS), and CREST (No.\,JPMJCR19T5) from Japan Science and Technology (JST). S.D.W. and B.R.O. acknowledge support via the UC Santa Barbara NSF Quantum Foundry funded via the Q-AMASE-i program under award DMR-1906325.

\noindent{\bf Author contributions:}
K.H. and T.S. conceived the project. M.R., K.I., K.Ogawa, K.Okada, Y.M., K.H., and T.S. performed magnetic penetration depth measurements and analyzed the data. Y.T., K.Okada, K.M., and Y.U. carried out high-pressure measurements. M.R. and S.L. performed electrical transport and X-ray diffraction measurements. R.G. and M.K. conducted electron irradiation experiments. B.R.O. and S.D.W. carried out sample preparation. M.R., K.H., and T.S. prepared the manuscript with inputs from R.G., M.K., B.R.O., and S.D.W. All authors discussed the experimental results.

\noindent{\bf Competing interests:} The authors declare no competing interests.

\noindent{\bf Additional information}

\noindent{\bf Supplementary Information} is available for this paper.

\noindent{\bf Correspondence and requests for materials} should be addressed to K.H. and T.S.

\bibliography{ref.bib}

\clearpage

\begin{figure*}[tbp]
    \includegraphics[width=\linewidth,]{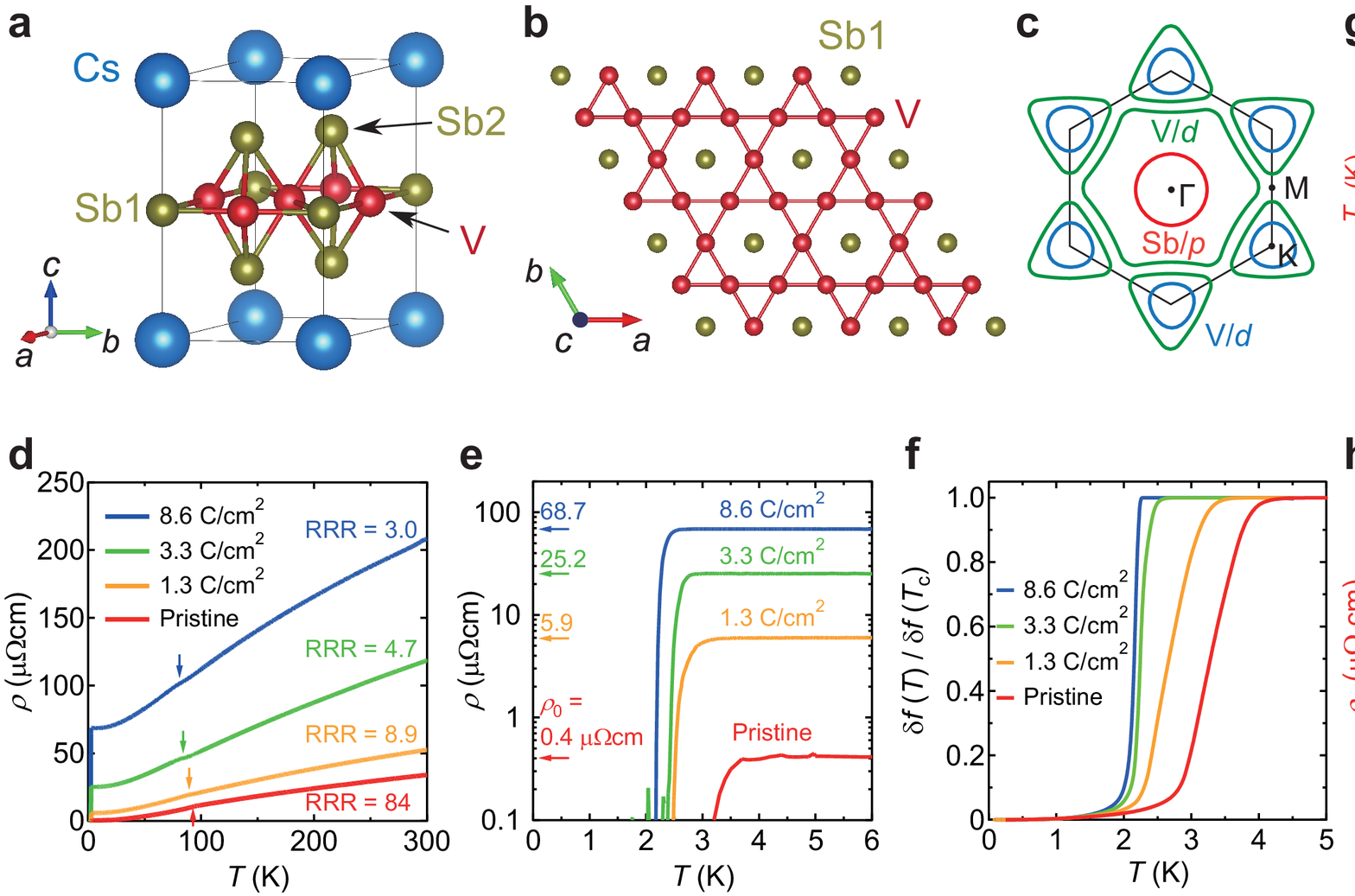}
\end{figure*}
\begin{figure*}[tbp]
    \caption{
{\bf Crystal structure, Fermi surface, and electron irradiation effects on the charge-density-wave and superconducting transition temperatures in CsV${_3}$Sb${_5}$.} {\bf a}, Crystal structure of CsV$_3$Sb$_5$. {\bf b}, V-Sb1 plane viewed from the $c$-axis direction. Whereas the V atoms form a two-dimensional kagome network, the Sb1 atoms are located at the hexagonal centers. {\bf c}, Schematic of the 2D Fermi surface of CsV$_3$Sb$_5$. The circular (red) and hexagonal (green) Fermi surfaces around the $\Gamma$ point are composed of the Sb $p$-orbitals and V $d$-orbitals, respectively, while the two triangular Fermi surfaces (blue and green) around the K point are formed by the V $d$-orbitals. {\bf d}, Temperature dependence of resistivity $\rho(T)$ in CsV${_3}$Sb${_5}$ single crystals with irradiated doses of 0 (pristine, red), 1.3 (orange), 3.3 (green), and 8.6 (blue) C/cm$^2$. The RRR values for each sample are listed. Arrows indicate the CDW transition temperatures determined from the temperature derivative of $\rho(T)$ (see Fig.\,4{\bf b},{\bf f},{\bf j}). Note that the $\rho(T)$ curves for the irradiated samples do not shift parallel to that of the pristine sample, which can be naturally understood by considering that \CVS\, is a multiband system (see Supplementary Information Sec.\,III). {\bf e}, Low-temperature resistivity below $6\,$K on a logarithmic scale. Arrows indicate the residual resistivity $\rho_0$ for each irradiated sample. {\bf f}, Temperature dependence of normalized frequency shifts of the TDO for each sample. {\bf g}, Superconducting and CDW transition temperatures $T_{\rm c}$ (left axis) and $T^{\ast}$ (right axis) as a function of irradiation dose. $T_{\rm c}$ is defined as the temperature at which the resistivity becomes zero (filled red circles), and the superfluid density becomes finite (open red circles). $T^{\ast}$ is defined as the temperature at which the derivative $d\rho/dT$ shows an abrupt change or dip (filled blue squares). {\bf h}, $\rho_0$ (left axis) and $\lambda (0)$ (right axis) as a function of irradiation dose. For the estimation of $\lambda(0)$, see Supplementary Information Sec.\,$\mathrm{IV}$.}
\label{F1}
\end{figure*}

\begin{figure*}[tbp]
    \includegraphics[width=0.9\linewidth]{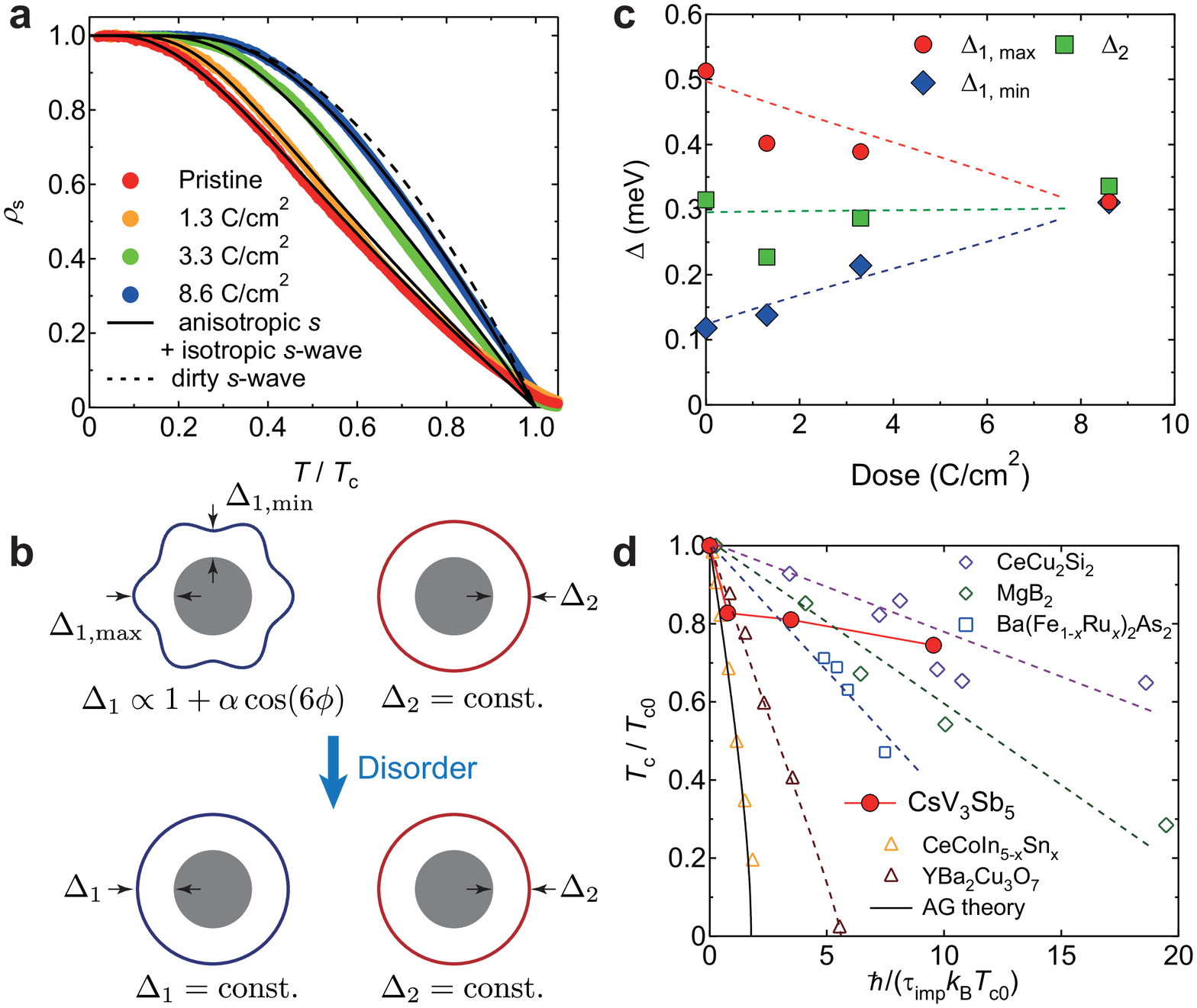}
\end{figure*}
\begin{figure*}[tbp]
    \caption{
{\bf Pair-breaking effect in CsV$_3$Sb$_5$.} {\bf a}, Temperature dependence of normalized superfluid density $\rho_{\rm s}(T) \equiv  \lambda^2 (0)/\lambda^2 (T)$ for the pristine (red), 1.3 (orange), 3.3 (green), and 8.6 (blue) C/cm$^2$ irradiated samples. Black solid lines are the fitting curves of the multigap model (clean limit). The black dashed line is the curve for the dirty $s$-wave (single gap) case. Note that $\rho_{\rm s}(T)$ for the 8.6 C/cm$^2$ irradiated sample with the relatively large value of $\xi/l = 2.3$ approaches the dirty limit (see Supplementary Information Sec.\,$\mathrm{V}$). {\bf b}, Schematic picture of the change in the superconducting gap structure against disorder. The introduction of electron irradiation changes the anisotropic superconducting gap structure to an isotropic one. {\bf c}, Gap sizes obtained from the fitting analysis of the superfluid density as a function of dose. Red circles and blue diamonds represent the maximum and minimum values of the anisotropic gap, $\Delta_{\rm{1,max}}$ and $\Delta_{\rm{1,min}}$, respectively. Green squares represent the gap values of the isotropic gap $\Delta_2$. Dotted lines are guides for the eyes. {\bf d}, Suppression of $T_{\rm c}$ in CsV$_3$Sb$_5$ (red circles) as a function of pair breaking parameter $g=\hbar/(\tau_{\rm imp}k_{\rm B}T_{\rm c0})$. For comparison, the results for Sn-substituted CeCoIn$_5$ (yellow triangles)\,\cite{115_impurity} and electron-irradiated YBa$_3$C$_3$O$_7$ (brown triangles)\,\cite{YBCO_e_irr} are plotted as examples of $d$-wave superconductors. Black solid line represents the suppression of $T_{\rm c}$ expected in the Abrikosov-Gor'kov (AG) theory. As an example of $s_{\pm}$-wave superconductors, the result of Ba(Fe$_{0.76}$Ru$_{0.24}$)$_2$As$_2$ (blue squares)\,\cite{BaRu122_irr_prozorov} is plotted. Also, the results of neutron-irradiated MgB$_2$ (green diamonds)\,\cite{MgB2_neutron_irr} and electron-irradiated CeCu$_2$Si$_2$ (purple diamonds)\,\cite{CCS_irr_yamashita} are plotted as examples of multigap $s$-wave superconductors. Dotted lines are guides for the eyes.}
    \label{F3}
\end{figure*}

\end{document}